**Simulation-based Algorithm for Determining Best Package Delivery Alternatives under Three Criteria: Time, Cost and Sustainability**

Suchithra Rajendran[a,b,*] and Aidan Harper[a]

[a]Department of Industrial and Manufacturing Systems Engineering, University of Missouri Columbia, MO 65211, USA
[b]Department of Marketing, University of Missouri Columbia, MO 65211, USA

[*]Corresponding Author:
Suchithra Rajendran
E-mail address: RajendranS@missouri.edu
Telephone: 573-882-7421

**Abstract**

With the significant rise in demand for same-day instant deliveries, several courier services are exploring alternatives to transport packages in a cost- and time-effective, as well as, sustainable manner. Motivated by a real-life case study, this paper focuses on developing a simulation algorithm that assists same-day package delivery companies to serve customers instantly. The proposed recommender system provides the best solution with respect to three criteria: cost, time, and sustainability, considering the variation in travel time and cost parameters. The decision support tool provides recommendations on the best alternative for transporting products based on factors, such as source and destination locations, time of the day, package weight, and volume. Besides considering existing new technologies like electric-assisted cargo bikes, we also analyze the impact of emerging methods of deliveries, such as robots and air taxis. Finally, this paper also considers the best delivery alternative during the presence of a pandemic, such as COVID-19. For the purpose of illustrating our approach, we consider the delivery options in New York City. We believe that the proposed tool is the first to provide solutions to courier companies considering evolving modes of transportation and under logistics disruptions due to pandemic.

**1. Introduction**

In the past decade, there has been a significant growth in same-day delivery (SDD) orders, specifically with the increasing presence of the online retail market. Roughly 51% of retailers offer SDD services, and the number is estimated to rise to 65% in the forthcoming years (Hartman, 2018). Same-day delivery has become a priority for retailers as a survey revealed that 75% of consumers would not even shop at a retailer if they could not deliver their products rapidly (Abdallah & Veeraraghavan, 2019). Moreover, another study concluded that nearly 88% of consumers are willing to pay an additional fee for same-day or faster delivery (PricewaterhouseCoopers, 2019). The concept of SDD is not only prevalent among retailers but is widespread among other service domains as well. According to Boyer et al. (2009), several businesses, such as office supply stores (e.g., Staples, Office Max, Office Depot), grocers (e.g., FreshDirect, Ocado), pharmaceutical and package deliveries (e.g., UPS, Airborne, FedEx, and Deutsche Post) offer direct customer deliveries.

There are several advantages and challenges associated with businesses as a result of incorporating the concept of SDD. One of the main benefits is the increased productivity level within the organization (Hartman, 2018). In particular, the warehouse efficiency is expected to rise substantially as a result of increased inventory turnover (i.e., the rate at which the products are moving in and out of the warehouses). On the other hand, as the need for faster delivery methods increases continuously, the shipping costs to fulfill customer orders will inflate as well (Abdallah & Veeraraghavan, 2019). For instance, in 2018 alone, shipping costs consumed $27.7 billion (12%) of Amazon's total revenues. With the growing number of online orders, companies also face the pressure to supply consumers with timely and reliable product distribution (Barclays, 2014; Devari et al., 2017).

Aside from large, traditional retailers that offer a vast amount of SDD services, other non-traditional courier companies also attempt to provide instant shipment services. For example, Instacart, Inc operates its organization without a warehouse or full-time shippers; rather, it utilizes independent contractors to deliver groceries to its customers from retailers (Kung & Zhong, 2016). Also, food delivery services like Postmates, UberEats, and Door Dash are similar to Instacart in hiring independent contractors to perform their SDD services. With this food delivery service, customers order food from their desired location, and an independent contractor will pick up the ordered food and deliver it for a service fee (Stonehem, 2016). In this process, the customers pay a flat rate for the meal they chose to order in addition to dynamic pricing associated with the independent contractor's time spent in picking up and delivering the food (Stonehem, 2016).

Traditional modes of vehicular transportation like trucks have a large vehicle capacity but are slower to deliver products due to traffic (Rajendran et al., 2015; Smith and Srinivas, 2019). Consequently, many companies and organizations have now started to leverage new technologies as well as adopting alternate methods of commutes to perform instant delivery obligations. For example, Amazon Inc. is currently capable of offering same-day and next-day delivery to 72% of the US population due to its immersive delivery network that has been culminating over the last four years (Kim, 2019). Amazon has recently introduced PrimeAir that can deliver packages in thirty minutes or less that are under five pounds using aerial vehicles (Amazon PrimeAir, 2020). The company has also received a patent for transforming public transportation into mobile storage stations, in which a removable delivery module will be attached to the vehicle (Baron, 2019). Similarly, FedEx is in the process of beta testing their robot named "Roxo," the FedEx SameDay Bot, in an attempt to curb the bottleneck associated with the last-mile delivery process. Roxo is designed to travel alongside roadways and sidewalks to deliver packages to residences and businesses (FedEx, 2020). Electric-Assisted (EA) cargo bikes are also being used to make deliveries in congested urban markets that provide a sustainable alternative to transporting goods (Fuldauer, 2019). EA cargo bikes are most cost-effective when they are in very close proximity to the delivery destination. A summary of the potential delivery methods for instant package shipments is presented in Table 1.

**Table 1:** Summary of the Current and Emerging Delivery Methods

| Delivery Method | Advantages | Disadvantages |
|---|---|---|
| **Emerging Method of Delivery** | | |
| 1. Aviation Vehicle | <ul><li>Packages can be delivered in thirty minutes or less on the same day as the order is processed.</li><li>Avoid traffic or congestion in heavily populated urban areas.</li></ul> | <ul><li>Not a feasible alternative for delivering a high volume and heavy packages.</li><li>Not as effective when dispatched to areas far from the computing site due to the small battery life.</li></ul> |

|  |  | • Smaller carbon footprint. |  |
|---|---|---|---|
| 2. Robot | | • SDD robot features pedestrian-safe technology plus other advanced technology like LiDAR and multiple cameras that allow it to travel safely while also being able to deliver the package in an efficient manner.<br>• Designed to travel alongside roadways and sidewalks to deliver packages to residences and businesses.<br>• Reduces carbon emissions of traditional channels of delivery. | • Not effective for rural areas with long-distance deliveries.<br>• Small payload that limits the weight and size of packages delivered.<br>• Since it is traveling alongside roadways and on sidewalks, it can take a considerable amount of time to deliver packages.<br>• Public safety could be put at risk due to these robots traveling around everywhere. |
| **Existing Method of Delivery** | | | |
| 3. EA Cargo Bike | | • Cost effective for delivering packages in urban areas.<br>• Smaller carbon footprint compared to delivery trucks. | • Not appropriate for delivering extremely heavy packages.<br>• Not effective for very long-distance deliveries. |
| 4. Cars | | • Companies can get their online orders to consumers much faster during lean periods.<br>• Effective for delivering heavy packages | • In case if delivery drivers work on a contract basis, confidentiality can be compromised as a result of individuals delivering the product without loyalty to the brand.<br>• Does not completely eliminate carbon emissions. |
| 5. Subway railroad | | • Offer a very high capacity for riders.<br>• Separated from traffic and extreme weather conditions | • Operates in fixed routes<br>• Restricted transfer points |

In this study, we develop a decision support system for courier and delivery service companies to deliver packages instantaneously. The best alternative with respect to three criteria (cost, time, and sustainability) is recommended considering the variability in travel time and cost. The decision support tool provides solutions based on factors, such as source and destination locations, time of the day, package weight, and volume. Besides considering existing new technologies like electric-assisted cargo bikes, and traditional methods (e.g., cars, subways, buses and walking), we also analyze the impact of emerging modes, such as robots and air taxis, for delivering products.

The remaining paper is organized as follows. Section 2 provides a review of the literature. The system settings and the model algorithm are detailed in Section 3, while the description of the data under consideration is presented in Section 4. The results and the managerial implications are given in Section 5, and the conclusions are provided in Section 6.

## 2. Literature Review

Research on same-day delivery has gained significant attention in recent years. Dayarian et al. (2020) conceptualized a way to utilize urban air vehicles for same-day home delivery settings. The authors developed a vehicle routing problem with drone resupply to provide direct-to-consumer (D2C) deliveries effectively. Their study mentioned that by 2050, 70% of the world's population would live within a major city, resulting in higher population densities, thus making it easier to achieve economies of scale in providing efficient SDD services. Klapp et al. (2018) analyzed the SDD system using the Dynamic Dispatch Waves Problem (DDWP). Using the DDWP models, the system's operator can determine whether or not to dispatch a single-vehicle loaded with orders ready for service. The researchers concluded that their best policy minimized the average cost of an a priori policy by 9.1% and maximized the orders delivered. A real problem that exists for companies attempting to break into the SDD category is the ability to evolve as an e-commerce platform without compromising revenue (Ulmer, 2017; Kawa et al., 2018; Scherr et al., 2018; Yao et al., 2019).

Lin et al. (2018) evaluated the efficiency of on-demand same-day delivery models with regards to fuel cost, transportation time, and emission cost. Three models are presented: same-day delivery with a commercial fleet, hub-and-spoke, and crowdsourcing. The results indicated that the hub-and-spoke model was the most cost-effective of the models for commercial carrier distribution services, but does not align well in providing same-day delivery services. Conversely, crowdsourcing has provided an avenue for minimized same-day delivery costs. Ulmer et al. (2018) analyzed a system of combining traditional delivery vehicles with drones to augment same-day delivery operations. To evaluate the most efficient mode of delivery for a particular circumstance, the researchers created a policy function approximation based on geographical districting. Their study highlighted how geographic districting had a major effect on increasing the number of same-day deliveries.

Haag & Hu (2019) detailed the bottleneck issues that can arise from the increase of traffic and congestion that same-day delivery presents. Due to the increase in same-day deliveries and the presence of online shopping, the number of trucks on tolled crossings in New York City and within their five boroughs grew from 32.6 million in 2013 to an estimated 35.7 million in 2018 (increased 9.4%). Similarly, a study conducted by The Rensselaer Polytechnic Institute Center of Excellence for Sustainable Urban Freight Systems was able to conclude that the average number of daily deliveries to New York City residences tripled to over 1.1 million shipments from 2009 through 2017. In an attempt to reduce this traffic congestion, Ni et al. (2019) aimed to solve the SDD problem by crowdsourced shipping methods. The authors were able to use a rolling horizon framework to predict the demand and leveraged the mixed-integer linear programming model intending to minimize the cost performance measure.

As same-day delivery becomes increasingly prevalent, Voccia et al. (2019) introduced a multi-vehicle dynamic pickup model considering delivery time constraints. Their results revealed that as on-demand requests increase, they can be filled when time windows are spread evenly throughout the day as opposed to requests being higher at certain times compared to others. Utilizing survey results from Nanjing, Xi et al. (2020) researched how same-day delivery had an overarching impact on local store shopping. The authors used a quasi-longitudinal design to determine the attributes affecting consumers' online shopping choices through same-day delivery on five different shopping experiences: supermarkets, convenience stores, vegetable markets, fruit stores, and restaurants. Hausmann et al. (2014) also argued that consumers are willing to pay more for same-day delivery, but there is a ceiling at which the convenience becomes too expensive for the consumer.

## 2.1 Contribution to the Literature

The paper contributes to the existing literature in multiple ways. We believe that this study is the first to develop a recommendation tool for courier companies to provide insights on the best delivery alternative during different times of the day in metropolitan cities. The proposed approach evaluates the alternate methods of commute and identifies the best solution with respect to three criteria: cost, time, and sustainability. In addition, the variability in travel time and cost parameters are considered to make the algorithm more applicable in real-life. Traditional courier service companies leverage a single way of transportation, such as trucks, cars, or bikes (e.g., Dayarian and Savelsbergh, 2017; Voccia et al., 2019). We also deviate from this policy by considering a combination of delivery options (e.g., $car \rightarrow subway \rightarrow bike$). Moreover, we also analyze the integration of emerging technologies, such as robots and air taxis, for delivering products. Finally, this paper also considers the best delivery alternative during the presence of a pandemic, such as COVID-19.

## 3. System Settings

Each customer order $c_j$ is defined by a tuple $(\lambda_j^{or}, \varphi_j^{or}, \lambda_j^{de}, \varphi_j^{de}, \tau_j, \vartheta_j, \omega_j)$ where $(\lambda_j^{or}, \varphi_j^{or})$ pair represents the origin latitude and longitude, $(\lambda_j^{de}, \varphi_j^{de})$ denotes the dropoff coordinates, $\tau_j$ is the time of the day, $\vartheta_j$ and $\omega_j$ are the package volume and weight, respectively. Different methods of transportation as detailed in Figure 1 and Table 2 are used, and the best performing alternative with regards to three measures is then determined using the algorithm given in Section 3.3.

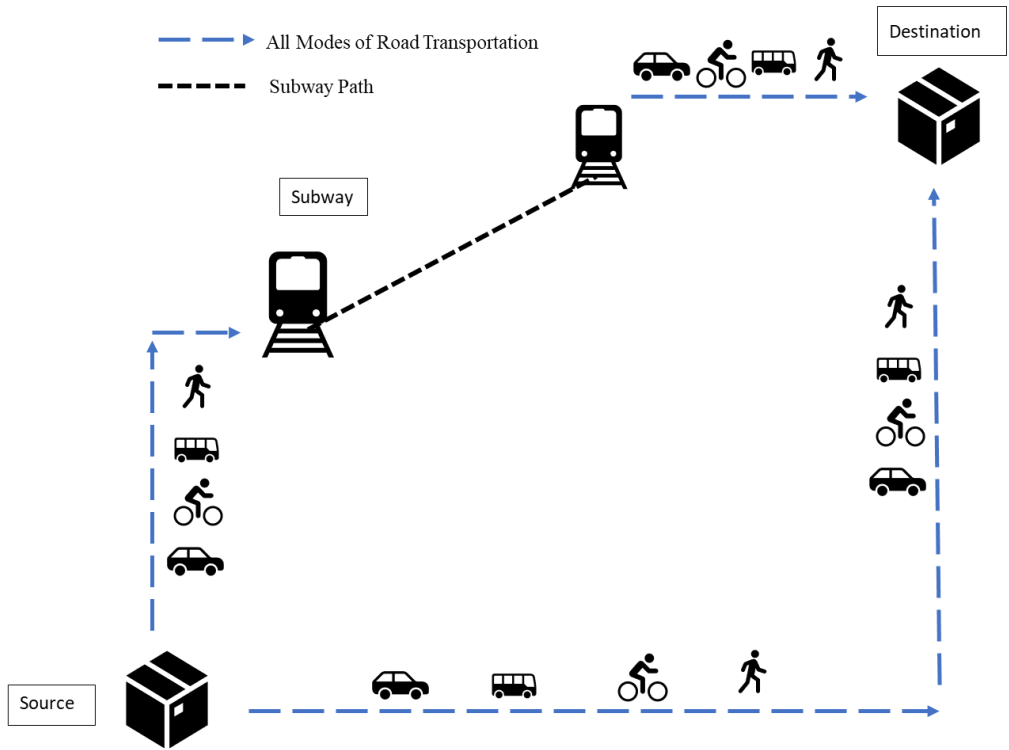

**Figure 1:** Overview of the Different Travel Methods

**Table 2:** Different Travel Alternatives Considered for Package Delivery

| Transportation Alternative $a$ | First-Mile | Middle-Mile | Last-Mile |
|:---:|:---:|:---:|:---:|
| 1 | Car | Car | Car |
| 2 | Bike | Bike | Bike |
| 3 | Walk | Walk | Walk |
| 4 | Bus | Bus | Bus |
| 5 | Car | Subway | Car |
| 6 | Car | Subway | Bike |
| 7 | Car | Subway | Bus |
| 8 | Car | Subway | Walk |
| 9 | Bike | Subway | Bike |
| 10 | Bike | Subway | Walk |
| 11 | Bike | Subway | Bus |
| 12 | Bike | Subway | Car |
| 13 | Walk | Subway | Walk |
| 14 | Walk | Subway | Bike |
| 15 | Walk | Subway | Bus |
| 16 | Walk | Subway | Car |
| 17 | Bus | Subway | Bus |
| 18 | Bus | Subway | Bike |
| 19 | Bus | Subway | Walk |
| 20 | Bus | Subway | Car |
| 21 | Car | Bus | Car |
| 22 | Car | Bus | Bike |
| 23 | Car | Bus | Bus |
| 24 | Car | Bus | Walk |
| 25 | Bike | Bus | Bike |
| 26 | Bike | Bus | Walk |
| 27 | Bike | Bus | Bus |
| 28 | Bike | Bus | Car |
| 29 | Walk | Bus | Walk |
| 30 | Walk | Bus | Bike |
| 31 | Walk | Bus | Bus |
| 32 | Walk | Bus | Car |
| 33 | Bus | Bus | Bike |
| 34 | Bus | Bus | Walk |
| 35 | Bus | Bus | Car |

As described in Table 2, each "mile" of delivery consists of various transportation options, including car, bus, subway, bike, and walk. In order to ensure that the most efficient delivery method is chosen, all combinations of transportation for each leg needs to be evaluated.

### 3.1 Model Notations

| Notation | Definition |
|---|---|
| $a$ | Index for an alternate travel option |
| $i, k, n$ | Index for first-, middle- and last-mile transportation methods, respectively |
| $j$ | Index for package order |
| $(\lambda_j^{or}, \varphi_j^{or})$ | Pair representing origin latitude and longitude |
| $(\lambda_j^{de}, \varphi_j^{de})$ | Pair representing destination latitude and longitude |
| $\tau_j$ | Time of the day during which the order $j$ is generated |
| $\vartheta_j$ | Package volume of order $j$ |
| $\omega_j$ | Package weight of order $j$ |
| $c^{su}$ | Subway ticket cost per ride |
| $c^{bu}$ | Bus ticket cost per ride |
| $c^{ca}$ | Operating cost of a car per mile (includes fuel cost and vehicle depreciation cost) |
| $c^{ea}$ | Operating cost of electric-assisted bike per mile (includes vehicle charging and depreciation cost) |
| $c^{pac}$ | Cost of parking a car per hour |
| $c^{pab}$ | Cost of parking a bike per hour |
| $c^{pe}$ | Delivery person wage per hour |
| $CO_2^{su}$ | $CO_2$ emission of subway per passenger mile |
| $CO_2^{bu}$ | $CO_2$ emission of the bus per passenger mile |
| $CO_2^{ca}$ | $CO_2$ emission of the car per passenger mile |
| $t_{i,j}^{fi}$ | First-mile travel time for order $j$ using transportation method $i$ (hours) |
| $t_{i,j}^{mi}$ | Middle-mile travel time for order $j$ using transportation method $i$ (hours) |

| Symbol | Description |
|---|---|
| $t_{i,j}^{la}$ | Last-mile travel time for order $j$ using transportation method $i$ (hours) |
| $wt_{k,j}^{fi}$ | Waiting time of messenger for the first-mile public transport (hours) |
| $wt_{k,j}^{mi}$ | Waiting time of messenger for the middle-mile public transport (hours) |
| $wt_{k,j}^{la}$ | Waiting time of messenger for the last-mile public transport (hours) |
| $d_{i,j}^{fi}$ | First-mile travel distance for order $j$ using transportation method $i$ (miles) |
| $d_{i,j}^{mi}$ | Middle-mile travel distance for order $j$ using transportation method $i$ (miles) |
| $d_{i,j}^{la}$ | Last-mile travel distance for order $j$ using transportation method $i$ (miles) |
| $o_{a,j}^{tr}$ | Overall travel time for order $j$ under alternative $a$ (hours) |
| $o_{a,j}^{co}$ | Overall cost for order $j$ in alternative $a$ |
| $o_{a,j}^{em}$ | Overall $CO_2$ emission for order $j$ under alternative $a$ |
| $o_{a,j}^{di}$ | Overall travel distance for order $j$ under alternative $a$ (miles) |
| $\delta_{i,j}^{fi}$ | 1 if transportation method $i$ is chosen for first-mile delivery for order $j$; 0 otherwise |
| $\delta_{k,j}^{mi}$ | 1 if transportation method $k$ is chosen for middle-mile delivery for order $j$; 0 otherwise |
| $\delta_{n,j}^{la}$ | 1 if transportation method $n$ is chosen for last-mile delivery for order $j$; 0 otherwise |
| $s_j^{or}$ | Closest subway station to the origin point $or$ of order $j$ |
| $s_j^{de}$ | Closest subway station to the destination point $de$ of order $j$ |

### 3.2 Model Assumptions

- Since the focus of the paper is on instant package delivery services, we assume that only one package order will be made per delivery, and multiple orders cannot be combined for proposing shipment alternatives.
- Package is picked up at the origin location $(\lambda_j^{or}, \varphi_j^{or})$ and is delivered at the customer specified destination location $(\lambda_j^{de}, \varphi_j^{de})$. Parcel will not be delivered in any in-between dropoff points.

- Package weight ($\omega_j$) is in-between 1 and 350 lbs, with over 50% of the weight being less than 5lbs. The impact of this distribution is analyzed in the sensitivity analysis section.
- Speed of biking and walking is influenced by the weight of the package.
- Speed of the bus and car varies based on the time of the day, and pickup and dropoff zones.
- Speed of other modes of commute, such as subway remains the same throughout the day.
- The time for the delivery person to travel to the package pickup address is assumed to be zero.

### 3.3 Algorithm Description

Once the package order described in the system settings is received by the courier company, the cyber-physical system computes the cost, time, and $CO_2$ emission associated with the different alternate settings discussed in Table 2.

For alternatives involving car as the only mode of transportation (i.e., $a = 1$), the total travel time is given as the sum of labor and vehicle operating costs, as shown in Equation (1).

$$o^{co}_{a=1,j} = c^{pe} \times o^{tr}_{a=1,j} + c^{ca} \times o^{di}_{a=1,j} + c^{pac} \tag{1}$$

For alternatives involving electric-assisted bikes as the only mode of transportation (i.e., $a = 2$), the total travel time is given as the sum of labor and bike operating costs, as shown in Equation (2).

$$o^{co}_{a=2,j} = c^{pe} \times o^{tr}_{a=2,j} + c^{ea} \times o^{di}_{a=2,j} + c^{pab} \tag{2}$$

For alternatives involving walk as the only mode of transportation (i.e., $a = 3$), the total travel time is the labor cost, as shown in Equation (3).

$$o^{co}_{a=3,j} = c^{pe} \times o^{tr}_{a=3,j} \tag{3}$$

For alternatives involving bus as the only mode of transportation (i.e., $a = 4$), the total travel time is given as the sum of labor cost (that includes that travel time and waiting time for buses) and bus ticket costs, as presented in Equation (4).

$$o^{co}_{a=4,j} = c^{pe} \times \{o^{tr}_{a=4,j} + wt^{fi}_{i \in bus,j} + wt^{mi}_{k \in bus,j} + wt^{la}_{l \in bus,j}\} + c^{bu} \times o^{di}_{a=4,j} \tag{4}$$

For the alternatives involving the subway mode of transportation as the middle mile (i.e., $a \geq 5$), the algorithm uses the package origin address and finds the closest subway station location (denoted by $s_j^{or}$). In order to reach this subway station, the service provider could travel by either car, bike, bus, or walk. The travel from the origin point $(\lambda_j^{or}, \varphi_j^{or})$ to the subway station $(s_j^{or})$ is referred to as the "first mile". The first mile travel time for order $j$ using mode $i$ is denoted by $t_{i,j}^{fi}$. The closest subway station to the destination location is then calculated. The subway travel is referred to as the "middle mile" (denoted by $t_{i,k \in sub}^{mi}$), where mode $k$ refers to the subway in this case. Similar to the first leg, to reach the actual destination, the service provider could travel by either car, bike, bus, or walk. The travel from the subway station, $s_j^{de}$, to the actual destination point $(\lambda_j^{or}, \varphi_j^{or})$ is referred to as the "last mile". The last mile travel time for order $j$ using mode $n$ is denoted by $t_{n,j}^{la}$. Since walk and bike methods of commute involve direct human interaction with transporting the package, weight restrictions are imposed on alternatives involving these options. A similar approach is adopted when bus is used for the middle-mile delivery, with bus travel being denoted by $t_{i,k \in bus}^{mi}$.

The total travel time for order $j$ using modes $i, k,$ and $n$ for the first-, middle- and last-mile, respectively, for alternative $a$ listed in Table 1, is computed using Equation (5).

$$o_{a,j}^{tr} = t_{i,j}^{fi} + t_{k,j}^{mi} + t_{n,j}^{la} \qquad (5)$$

Similarly, the total travel cost for order $j$ using modes $i, k,$ and $n$ for the first-, middle- and last-mile, respectively, for alternative $a$ listed in Table 1 is computed using Equation (6). The total travel cost consists of three components: messenger labor cost, vehicle operating cost, and bus and subway ticket purchasing cost.

$$\begin{aligned}
o_{a\geq5,j}^{co} = & c^{pe} \times \{t_{i\in car,j}^{fi} + t_{i\in bus,j}^{fi} + t_{i\in walk,j}^{fi} + t_{i\in bike,j}^{fi}\} \\
& + c^{pe} \times \{t_{k\in sub,j}^{mi} + t_{k\in bus,j}^{mi}\} \\
& + c^{pe} \times \{t_{n\in car,j}^{la} + t_{n\in bus,j}^{la} + t_{n\in walk,j}^{la} + t_{n\in bike,j}^{la}\} \\
& + c^{pe} \times \{wt_{i\in sub,j}^{fi} + wt_{k\in bus,j}^{mi} + wt_{n\in bus,j}^{la}\} + c^{pe} \times wt_{k\in sub,j}^{mi} \\
& + c^{ea} \times \{d_{i\in bike,j}^{fi} + d_{n\in bike,j}^{la}\} + c^{ca} \times \{d_{i\in car,j}^{fi} + d_{n\in car,j}^{la}\} \\
& + c^{bu} \times \{\delta_{i\in bus,j}^{fi} + \delta_{k\in bus,j}^{mi} + \delta_{n\in bus,j}^{la}\} + c^{su} \times \delta_{k\in sub,j}^{mi} \\
& + c^{pac} \times \{\delta_{i\in car,j}^{fi} + \delta_{n\in car,j}^{la}\} + c^{pab} \times \{\delta_{i\in bike,j}^{fi} + \delta_{n\in bike,j}^{la}\}
\end{aligned} \quad (6)$$

The total $CO_2$ emission for order $j$ using modes $i, k,$ and $n$ for the first-, middle- and last-mile, respectively, for alternative $a$ listed in Table 1 is computed using Equation (7).

$$\begin{aligned}
o_{a,j}^{co} = & CO_2^{ca} \times \{d_{i\in car,j}^{fi} + d_{k\in car,j}^{mi} + d_{n\in car,j}^{la}\} \\
& + CO_2^{bu} \times \{d_{i\in bus,j}^{fi} + d_{k\in bus,j}^{mi} + d_{n\in bus,j}^{la}\} + CO_2^{su} \times d_{k\in sub,j}^{mi}
\end{aligned} \quad (7)$$

Based on the calculations, the algorithm determines the best alternative with regards to the three criteria: total delivery time, cost, and pounds of $CO_2$ emitted. The clients will receive these alternatives on their smart devices and can choose among the options according to their preference. After the package is safely delivered, an invoice will be created, and customers will be notified about the product delivery. The sequence of events in booking a package delivery order is illustrated in Figure 2.

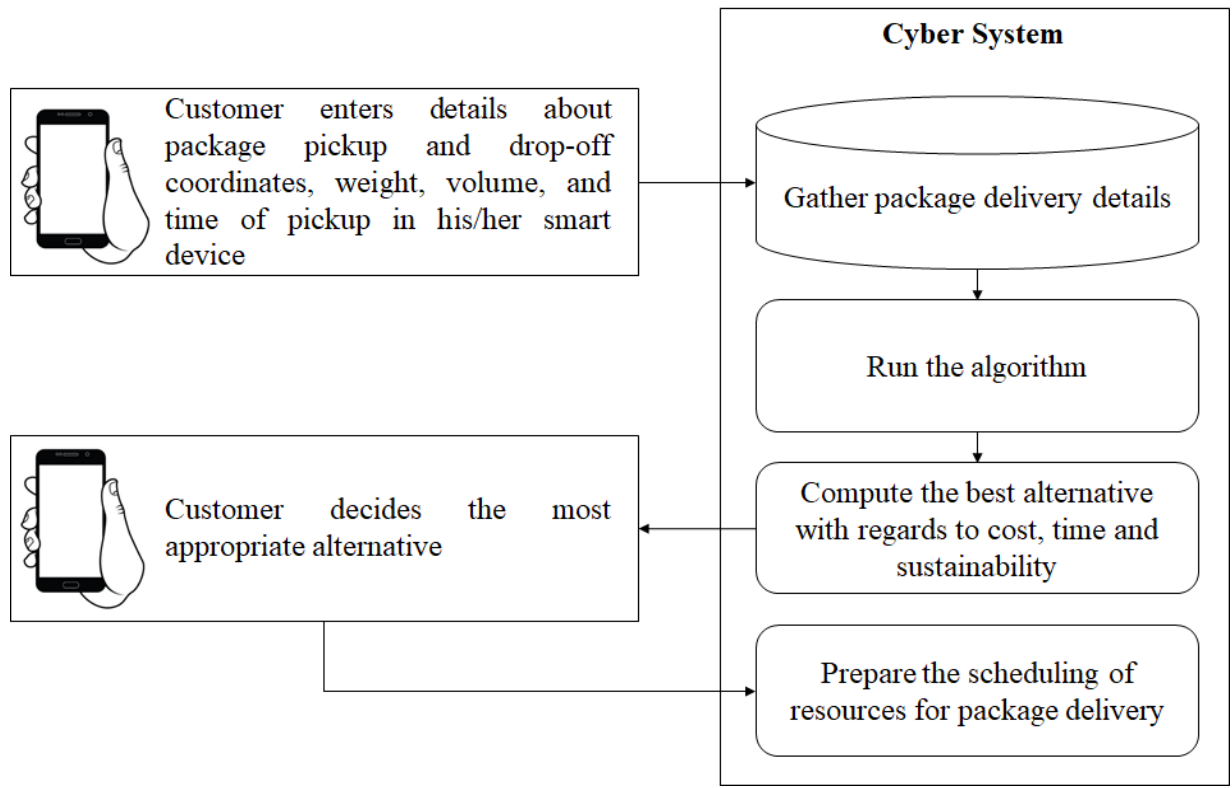

**Figure 2:** Sequence of Events in Booking a Package Delivery Order

## 4. Data Description

For the purpose of illustration, the algorithm described in Section 3.3 is applied to New York City (NYC). NYC is chosen due to two reasons. This city has a very high population density, with over 27,000 people per square mile (NYC Department of City Planning, 2020). This massive population drives the demand for SDD, whether it be retail, food, or pharmaceuticals. Next, more than two million deliveries are made in NYC every day, with nearly a million freight trips via New York's five boroughs (Cheng, 2019). In order to develop a recommendation tool for courier companies in NYC, we partition the urban region into 13 zones: Lower Manhattan, East Village, Midtown Manhattan, Upper East Side, Hell's Kitchen, Upper West Side, Harlem, Upper Manhattan, Sugar Hill, Washington Heights, Hudson Heights, Fort George, and Inwood. Based on the boundaries of each zone, with the GPS coordinates of the order source and destination, each package is mapped to the corresponding start and end zones. With the information provided in New York City (2006), Mazur (2015), and New York City Department of Transportation (2018), Salama and Srinivas (2020), and the US Department of Transportation (2020), Table 3 summarizes the various parameters obtained for this study. The variation in driver wage and travel time parameters are also evaluated in the proposed model. Based on the information on the delivery driver salary obtained from Glassdoor (2021), it is sampled using a normal distribution with a mean of $13.37 and standard deviation of $3 (i.e., messenger salary $\approx N(13.37,3)$). The travel time variability of cars and buses is estimated from Yazici et al. (2017), and the variation in subway travel time is obtained from Hu (2018). For the sake of simplicity, the variations in the bike and walking speeds are assumed to be 10% of their corresponding mean value, following a uniform distribution.

**Table 3:** Summary of the Model Parameters

| Parameter | Transportation Method | Value |
| --- | --- | --- |
| Speed (average, standard deviation) (mph) | Bike | (5, 0.5) |
| | Walking | (3.5, 0.35) |
| | Subway | (17, 8) |
| | Cars and busses | (7.1, 1.3) |
| $CO_2$ Emission (per passenger mile) | Bus | 0.64 pounds |
| | Car | 0.96 pounds |
| | Subway | 0.33 pounds |
| Median delivery messenger salary | | $13.37 per hour |
| Weight range | | 50% of the weight < 5 lbs |

## 5. Results

This section presents the results of the base case setting. Besides, the impact of emerging methods of transportation, such as robots and air taxis, for delivering products is analyzed. Finally, the best delivery alternatives during the presence of a pandemic, such as COVID-19, are also examined in this section.

## 5.1 Base Case

As mentioned earlier, the proposed approach evaluates the alternate methods of commute and identifies the best solution with respect to three criteria: cost, time, and sustainability. The simulation model is sampled for one million orders with diversified pickup and dropoff points, as well as, order times. The origin and destination locations are sampled considering the population density of different regions. In other words, densely-populated regions, such as Midtown Manhattan, will ship and receive more packages than less-crowded zones, such as Fort George.

Tables 4 and 5 show the best time- and cost-effective alternatives for different distance bins. On analyzing the best time setting, one can observe that cars and bikes are widely being used for short commutes. More specifically, cars are leveraged during early mornings and late nights, which is perhaps due to the relatively less traffic, while packages are expected to be delivered by bikes during rush hours. Alternatives involving buses and subways are not as significant as the other modes of commutes, primarily due to the additional time incurred as a result of parking the bike/car and then waiting for these public transportation methods. Whereas as the distance increases between two and four miles, $bike$ and $walk \rightarrow subway \rightarrow walk$ are observed to be the best alternatives. This might be due to the reason that most of the packages under the medium distance category are transported between the east and west side of the town. These are well-connected by subways, and hence, the delivery person could easily walk to the nearest metro for the first- and last-mile delivery.

During rush hours, the speed of bikes will be less than usual, especially in tourism regions, such as the Time Square and the Empire State building, despite the presence of bike lanes. For distance more than six miles, car is the best alternative. This is expected because most of the long-distance packages are sent between the north and south ends of the city, where the public transport connectivity is relatively low. Moreover, the traffic in these ends of the city is comparatively less compared to midtown that includes several tourist destinations. For all the distance settings, buses and subways operate less frequently during late nights and early mornings, and hence they are not the most effective method of delivery during these times.

**Table 4:** Best Time-Effective Alternatives for Different Times of the Day

| Distance in Miles | Best Time Option | | | |
| --- | --- | --- | --- | --- |
| | Early Morning | Morning & Early Afternoon | Late Afternoon & Evening | Night |
| < 1 | car → car → car | bike → bike → bike | bike → bike → bike | car → car → car |
| 1 – 2 | car → car → car | bike → bike → bike | bike → bike → bike | car → car → car |
| 2 – 3 | car → car → car | walk → subway → walk | walk → subway → walk | car → car → car |
| 3 – 4 | car → car → car | walk → subway → walk | walk → subway → walk | car → car → car |
| 4 – 5 | car → car → car | bike → subway → bike | bike → subway → bike | car → car → car |
| 5 – 6 | car → car → car | bike → subway → bike | bike → subway → bike | car → car → car |
| 6 – 7 | car → car → car | car → car → car | car → car → car | car → car → car |
| > 7 | car → car → car | car → car → car | car → car → car | car → car → car |

**Table 5:** Best Cost-Effective Alternatives for Different Times of the Day

| Distance in Miles | Best Cost Option | | | |
| --- | --- | --- | --- | --- |
| | Early Morning | Morning & Early Afternoon | Late Afternoon & Evening | Night |
| < 1 | bike → bike → bike | bike → bike → bike | bike → bike → bike | bike → bike → bike |
| 1 – 2 | bike → bike → bike | bike → bike → bike | bike → bike → bike | bike → bike → bike |
| 2 – 3 | bike → bike → bike | bike → bike → bike | bike → bike → bike | car → car → car |
| 3 – 4 | car → car → car | walk → subway → walk | walk → subway → walk | car → car → car |
| 4 – 5 | car → car → car | bike → subway → bike | bike → subway → bike | car → car → car |
| 5 – 6 | car → car → car | bike → subway → bike | bike → subway → bike | car → car → car |
| 6 – 7 | car → car → car | car → car → car | car → car → car | car → car → car |
| > 7 | car → car → car | car → car → car | car → car → car | car → car → car |

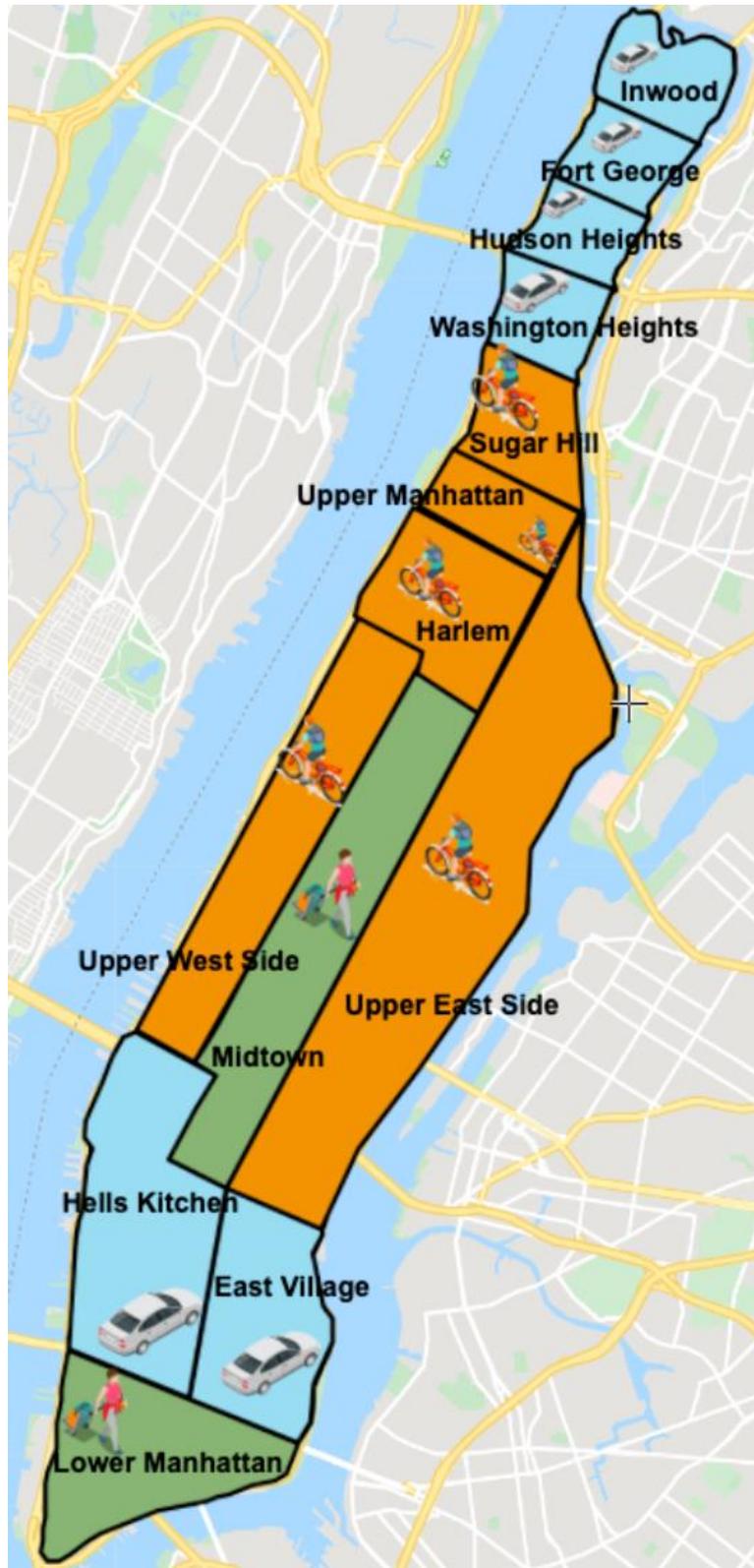

**Figure 3:** Best Time-Effective Alternative for First-/Last-Mile Deliveries

With regard to the cost-effective option (Table 5), we can notice that bikes are being widely suggested by our analysis as the most economical method, given the lack of vehicle operating costs (unlike in the case of a car) or ticket purchasing cost (in case of the bus and subways). For medium distances, we can notice that bikes are not recommended for the entire commute because of the tradeoff that has to be made between subway/bus purchasing cost and labor cost associated with the product delivery and waiting for public transportation. For the most sustainable option, we find that *bike* and *walk* constitute over 99% of the results, since a bike can carry up to 300 lbs of weight. Figure 3 shows the quickest first-/last-mile transportation alternative across different zones of NYC. We can observe that bike and walk are dominating the midtown regions of NYC. This is logical as midtown includes Central Park, where there is more access for bikes and less for buses, which brings down the average speed of vehicles in that zone. The alternatives in the upper portion of Manhattan varies with different zones, while it is widely recommended to use car due to the relatively fewer subway stations and trains frequency.

### 5.2 Sensitivity Analysis

In this section, the impact of several parameters on the performance measures is analyzed. Section 5.2.1 examines the influence of the distribution of delivery weights. Section 5.2.2 discusses the effect of the presence of a pandemic, while Section 5.2.3 investigates the future modes of delivery options, namely robots and air taxis.

### 5.2.1 Impact of Delivery Weight Distribution

In the base case setting, we assume that 50% of the packages weigh less than 5 lbs. We vary the weight distribution to identify its impact on the model results. In weight setting-1 (WS1), 15% of the total weight is less than 5 lbs, while 35% of the total weight is assumed to be less than 5 lbs in WS-2. WS-3 is assumed to be the base case, and 85% of the weight is less than 5 lbs in WS-4. From the results presented for the time-effective method (Table 6), we can see that for settings WS-1 and WS-2, buses and cars are significantly used (especially for very short and very long commutes). For all settings, walk appears to be a good first- and last-mile option for medium distance commutes. This might be due to the reason that most of the packages under the medium distance category are transported between the east and west side of the town, and these regions are well-connected by buses and subways. For all the weight distribution settings, buses are not used for long-distance commutes due to the direct relationship between the distance (from the source to destination) and the number of stops.

For WS-4, bikes and walks are the best methods of commutes across almost all settings. This is perhaps because the bike speed is inversely proportional to the package weight and WS-4 setting has 85% of the weight under 5 lbs. As in the other cases, for sustainability, the *bike* option is again the most sustainable across most weight intervals. Table 7 shows the best alternative when the cost has to be minimized. We can see that the recommendations for settings WS-3 and WS-4 are the same. Whereas since settings WS-1 and WS-2 involve heavy packages, bikes are not preferred, and buses and cars are predominantly used. For WS-1 and WS-2, despite the presence of vehicle operating cost, cars are the preferred mode of transportation during early morning and late night, due to reduced traffic. During the course of the day, as a result of the increase in traffic, the model recommends availing of the bus and subway services.

**5.2.2 Presence of Pandemic**

During the presence of a pandemic, such as COVID-19, same-day delivery options have to quickly adapt to their environment to assure that their operations are being conducted in a manner that does not distort their efficiency. For instance, on-demand, same-day third-party grocery delivery services like Instacart, Inc. have hired an additional 300,000 independent contractors to meet the demand of shopper needs (Burszytynsky, 2020). Due to social distance regulations, companies like Office Depot, Inc. are also introducing SDD of various office supplies to immediately cater to individuals who have been forced to work remotely (Office Depot, 2020). Also, services, such as same-day food delivery (e.g., Postmates, Uber Eats), have seen a surge in sales due to the contactless delivery option permitted by these corporations (Butler, 2020). This instantaneous delivery service has particularly become essential during pandemics.

**Table 6:** Best Time Alternatives for Different Weight Settings

| Scenario | WS-1 (15% <5lbs) | WS-2 (35% <5lbs) | WS-3 (50% <5lbs) | WS-4 (85% <5lbs) |
|---|---|---|---|---|
| **Distance** | | | | |
| < 1 | car → car → car | car → car → car | bike → bike → bike | bike → bike → bike |
| 1 – 2 | car → car → car | bus → subway → bus | bike → bike → bike | bike → bike → bike |
| 2 – 3 | bus → bus → bus | walk → bus → bus | walk → subway → walk | walk → subway → walk |
| 3 – 4 | bus → bus → bus | walk → bus → bus | walk → subway → walk | walk → subway → walk |
| 4 – 5 | bus → subway → bus | walk → bus → bus | bike → subway → bike | bike → subway → bike |
| 5 – 6 | bus → subway → bus | bus → subway → bus | bike → subway → bike | bike → subway → bike |
| 6 – 7 | car → subway → car | car → subway → car | car → car → car | car → car → car |
| > 7 | car → subway → car | car → subway → car | car → car → car | car → car → car |
| **Time of the day** | | | | |
| Early Morning | car → car → car | car → car → car | car → car → car | car → car → car |
| Morning & Early Afternoon | bus → subway → bus | walk → bus → bus | walk → subway → walk | walk → subway → walk |
| Late Afternoon & Evening | bus → subway → bus | Walk → bus → bus | walk → subway → walk | walk → subway → walk |
| Night | car → car → car | car → car → car | car → car → car | car → car → car |

Table 7: Best Cost Alternative for Different Weight Settings

| Scenario | WS-1 (15% <5lbs) | WS-2 (35% <5lbs) | WS-3 (50% <5lbs) | WS-4 (85% <5lbs) |
|---|---|---|---|---|
| **Distance** | | | | |
| < 1 | car → car → car | car → car → car | bike → bike → bike | bike → bike → bike |
| 1 – 2 | car → car → car | bus → subway → bus | bike → bike → bike | bike → bike → bike |
| 2 – 3 | bus → bus → bus | walk → bus → bus | bike → bike → bike | bike → bike → bike |
| 3 – 4 | bus → bus → bus | walk → bus → bus | walk → subway → walk | walk → subway → walk |
| 4 – 5 | bus → subway → bus | walk → bus → bus | bike → subway → bike | bike → subway → bike |
| 5 – 6 | bus → subway → bus | bus → subway → bus | bike → subway → bike | bike → subway → bike |
| 6 – 7 | car → subway → car | car → subway → car | car → car → car | car → car → car |
| > 7 | car → subway → car | car → subway → car | car → car → car | car → car → car |
| **Time of the day** | | | | |
| Early Morning | car → subway → car | car → subway → car | car → car → car | car → car → car |
| Morning & Early Afternoon | car → subway → bus | car → subway → bus | bike → subway → bike | bike → subway → bike |
| Late Afternoon & Evening | bus → subway → car | bus → subway → bus | bike → subway → bike | bike → subway → bike |
| Night | car → subway → car | car → subway → car | car → car → car | car → car → car |

**Table 8:** Best Time-Effective Alternatives During the Presence of a Pandemic

| Distance in Miles | Best Time Option | | | |
|---|---|---|---|---|
| | Early Morning | Morning & Early Afternoon | Late Afternoon & Evening | Night |
| < 1 | car → car → car | car → car → car | car → car → car | car → car → car |
| 1 – 2 | car → car → car | car → car → car | car → car → car | car → car → car |
| 2 – 3 | car → car → car | car → car → car | car → car → car | car → car → car |
| 3 – 4 | car → car → car | car → car → car | car → car → car | car → car → car |
| 4 – 5 | car → car → car | car → car → car | car → car → car | car → car → car |
| 5 – 6 | car → car → car | car → car → car | car → car → car | car → car → car |
| 6 – 7 | car → car → car | car → car → car | car → car → car | car → car → car |
| > 7 | car → car → car | car → car → car | car → car → car | car → car → car |

**Table 9:** Best Cost-Effective Alternatives During the Presence of a Pandemic

| Distance in Miles | Best Cost Option | | | |
|---|---|---|---|---|
| | Early Morning | Morning & Early Afternoon | Late Afternoon & Evening | Night |
| < 1 | bike → bike → bike | bike → bike → bike | bike → bike → bike | bike → bike → bike |
| 1 – 2 | bike → bike → bike | bike → bike → bike | bike → bike → bike | bike → bike → bike |
| 2 – 3 | bike → bike → bike | bike → bike → bike | bike → bike → bike | bike → bike → bike |
| 3 – 4 | car → car → car | walk → subway → walk | walk → subway → walk | car → car → car |
| 4 – 5 | car → car → car | bike → subway → bike | bike → subway → bike | car → car → car |
| 5 – 6 | car → car → car | car → car → car | car → car → car | car → car → car |
| 6 – 7 | car → car → car | car → car → car | car → car → car | car → car → car |
| > 7 | car → car → car | car → car → car | car → car → car | car → car → car |

Tables 8 and 9 show the results of the best alternatives for the cost and time settings. During a lockdown, due to the decrease in on-road traffic, the average vehicle speed increases from 13 mph to 52 mph (Hu, 2020). As a result, *car* travel is chosen to be the best delivery option with regards to time. Concerning cost-effective options, for distances less than 5 miles, bikes, subway, and walk mostly appear to be the best alternative. Nevertheless, for longer distance deliveries, the usage of cars is more economical. Also, since subways are less operational during late night and early mornings after the onset of this pandemic (Metropolitan Transportation Authority, 2020), car is reported to be the best alternative during those times, while mornings and afternoon deliveries avail of the subway services.

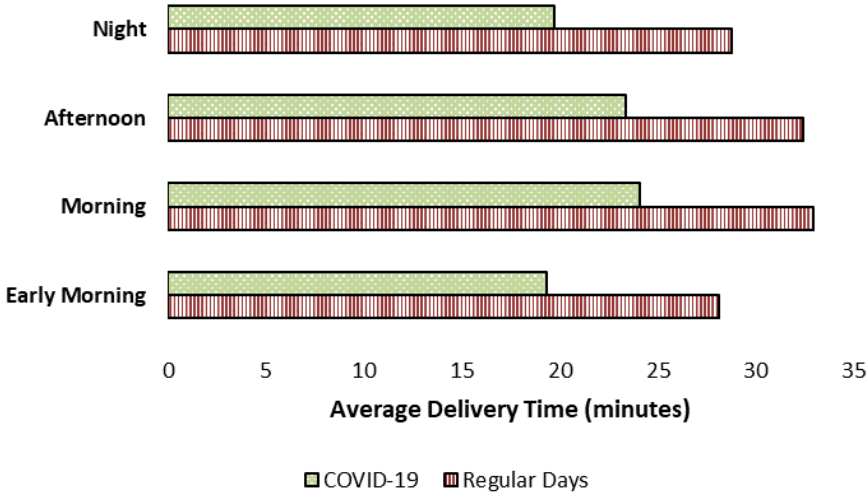

**Figure 4:** Comparison of Average Delivery Times for Different Times of the Day During the Presence of a Pandemic

Figure 4 shows the average delivery times relative to each time of the day. As shown, pre-COVID and during the pandemic have the same delivery time pattern, with afternoon and morning having the most extended delivery times as traffic is still increasing. Early morning and night-time deliveries are relatively less, as anticipated. Moreover, we can notice that the average delivery time during COVID-19 is approximately 10 minutes less than before the pandemic. Interestingly, from Figure 5, we can see that the percentage increase in the average delivery time appears to be almost linear for both pre-COVID and during the pandemic settings. Nevertheless, during the pandemic, due to the significant decrease in traffic, average time savings of more than 50% is achieved across different distance settings.

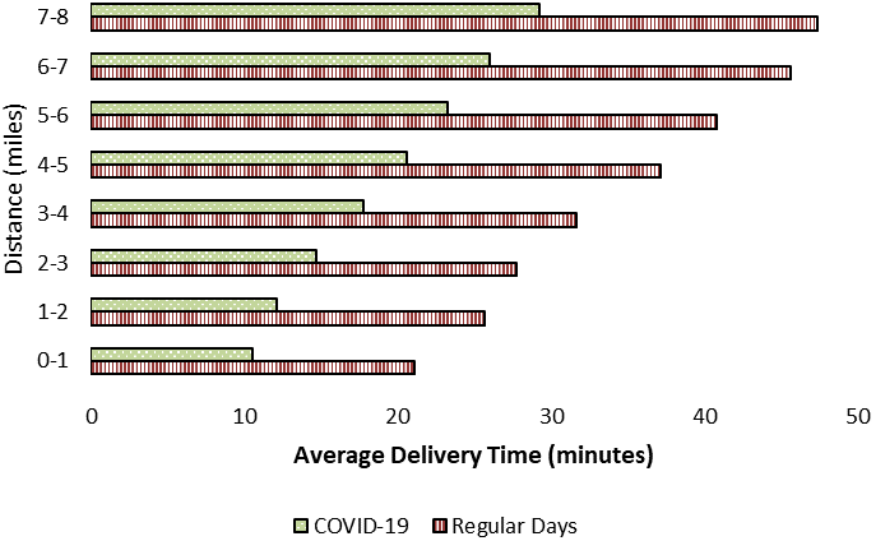

**Figure 5:** Comparison of Average Delivery Times vs. Distance During the Presence of a Pandemic

**5.2.3 Emerging Modes of Commute**

In recent years, with the increase in traffic and considering the environmental impact, many logistics companies have now started to leverage new technologies, as well as, adopting alternate methods of commutes to perform instant delivery obligations. In this section, we analyze the impact of robots and air taxis (urban air mobility services that are anticipated to be launched in metropolitan cities in the forthcoming years) on package delivery services (Rajendran et al., 2021). While robots are used for first- and last-mile services, air taxis are considered for middle-mile options. A schematic representation of the delivery alternatives using emerging technology is shown in Figure 6.

The parameter settings, such as the average robot weight carrying capacity, range, and speed, are obtained from prior studies (Starship Technologies, 2020; The Business and Technology of Enterprise AI, 2020), which are 20 lbs, 3.5 miles and four mph, respectively. For delivery orders involving heavy packages or longer first- and/or last-mile distances, this delivery method will be disregarded. The location of air taxi stations is obtained from Rajendran and Shulman (2020). Table 10 portrays the different combinations of methods of commutes considered for the analysis of emerging technology. Based on the results, it is observed that the alternatives listed in Table 10 do not perform well for the best cost, time, and sustainable options. Although air taxis operate at an average speed of 170 mph (Holden and Goel, 2016; Rajendran and Zack, 2019), there are only very few operational sites (in comparison to subway stations) within NYC, and hence, the majority of the commute has to be made on the road. Nevertheless, in the future, after the launch of air taxi services, with more infrastructure sites becoming operational (Rajendran and Srinivas, 2020), air taxis could serve as a potential middle-mile option with regards to the time criteria.

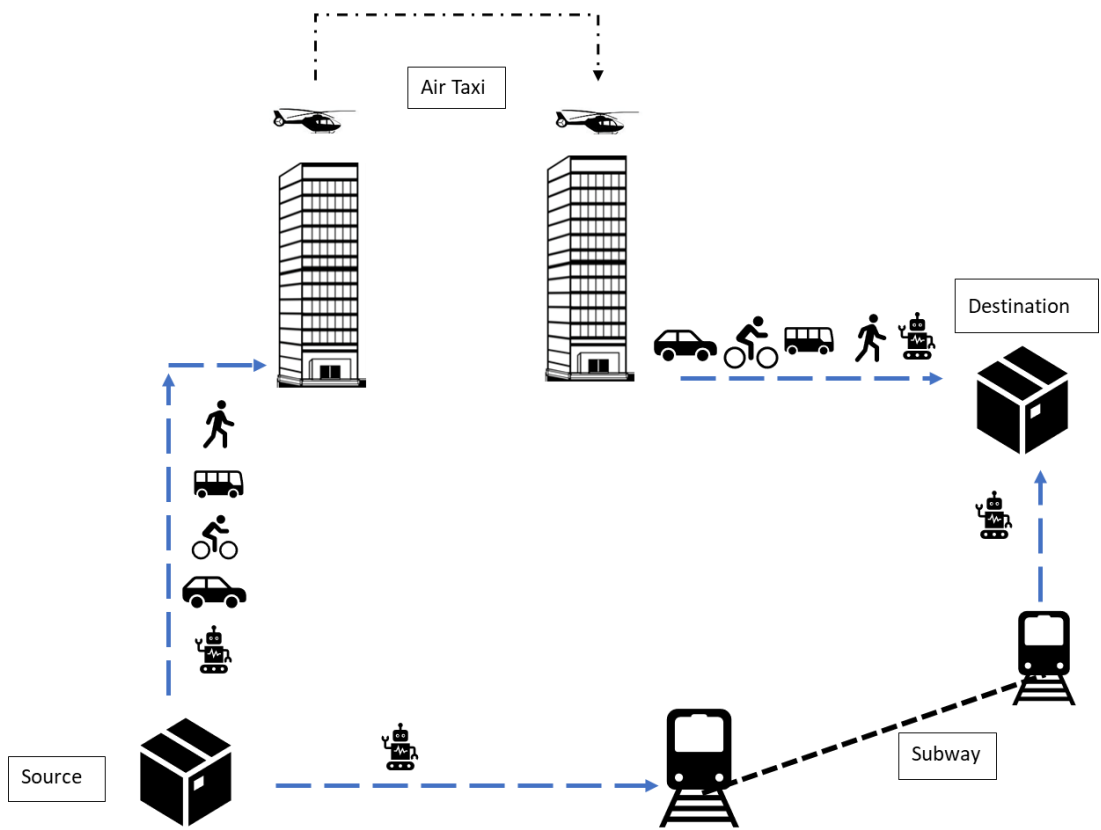

**Figure 6:** Overview of the Emerging Alternatives Considered

**Table 10:** Alternatives Considering Emerging Modes of Commute

| Transportation Option $a$ | First Mile | Middle Mile | Last Mile |
|---|---|---|---|
| 36 | Robot | Subway | Robot |
| 37 | Robot | Subway | Car |
| 38 | Robot | Subway | Bus |
| 39 | Robot | Subway | Bike |
| 40 | Robot | Subway | Walk |
| 41 | Car | Subway | Robot |
| 42 | Bus | Subway | Robot |
| 43 | Bike | Subway | Robot |
| 44 | Walk | Subway | Robot |
| 45 | Robot | Bus | Robot |
| 46 | Robot | Bus | Car |
| 47 | Robot | Bus | Bus |
| 48 | Robot | Bus | Bike |
| 49 | Robot | Bus | Walk |
| 50 | Car | Bus | Robot |
| 51 | Bus | Bus | Robot |
| 52 | Bike | Bus | Robot |
| 53 | Walk | Bus | Robot |
| 54 | Robot | Air taxi | Robot |
| 55 | Robot | Air taxi | Car |
| 56 | Robot | Air taxi | Bus |
| 57 | Robot | Air taxi | Bike |
| 58 | Robot | Air taxi | Walk |
| 59 | Car | Air taxi | Robot |
| 60 | Bus | Air taxi | Robot |
| 61 | Bike | Air taxi | Robot |
| 62 | Walk | Air taxi | Robot |
| 63 | Car | Air taxi | Car |
| 64 | Car | Air taxi | Bike |
| 65 | Car | Air taxi | Bus |
| 66 | Car | Air taxi | Walk |
| 67 | Bike | Air taxi | Bike |
| 68 | Bike | Air taxi | Walk |
| 69 | Bike | Air taxi | Bus |
| 70 | Bike | Air taxi | Car |
| 71 | Walk | Air taxi | Walk |
| 72 | Walk | Air taxi | Bike |
| 73 | Walk | Air taxi | Bus |
| 74 | Walk | Air taxi | Car |
| 75 | Bus | Air taxi | Bus |
| 76 | Bus | Air taxi | Bike |
| 77 | Bus | Air taxi | Walk |
| 78 | Bus | Air taxi | Car |

## 5.3 Managerial Recommendations

Based on our simulation study analysis, we propose the following managerial implications.

- For the quickest delivery alternative, it is widely recommended to use cars and bikes for short commutes, whereas $walk \rightarrow subway \rightarrow walk$ and $bike \rightarrow subway \rightarrow bike$ are suggested for medium distances (between 1 and 6 miles). Whereas for long distances, our algorithm recommends leveraging cars.

- We can notice that alternatives involving bikes are being widely suggested by our analysis as the most economical method, given the lack of vehicle operating cost (unlike in the case of a car) or ticket purchasing cost (unlike bus and subway).
- The *bike* alternative is not recommended for the entire medium-distance commutes because of the tradeoff made between subway purchasing cost and labor cost associated with the product delivery.
- For the most sustainable option, we find that bike and walk constitute over 99% of the results since a bike can carry up to 300 lbs of weight.
- In early mornings and nights, due to less traffic, car is the fastest alternate commute option. Whereas, during the morning and afternoon, with peak traffic, it is preferred to deliver packages via bike, subway and walk.
- With regards to the quickest first- and last-mile delivery alternatives, bike and walk are dominating the midtown regions of NYC, due to the existence of Central Park, where there is more access for bikes, and less for buses.
- The alternatives in the upper portion of Manhattan varies with different zones, while it is widely recommended to use car due to the relatively fewer subway stations and trains frequency.
- For settings involving heavier packages, car is predominantly used for very short and very long commutes for the quickest delivery option, whereas, with the increase in the percentage of low weight packages, alternatives involving bus and subway are preferred for several settings.
- During the time of a pandemic lockdown, as a result of the decrease in on-road traffic, car travel is recommended to be the best delivery option with regards to time. Concerning cost effectiveness, bikes and cars appear to be the best alternative for short- and long-distance deliveries, respectively. In contrast, $walk \rightarrow subway \rightarrow walk$ and $bike \rightarrow subway \rightarrow bike$ are more economical alternatives for medium distance commutes.
- Since subways and buses are less operational during late night and early mornings after the onset of this pandemic, car is reported to be the best alternative during those times, while mornings and afternoon deliveries avail of these public transportation services.
- The average delivery time during COVID-19 is approximately 10 minutes less than before the pandemic.

- During the pandemic, due to the significant decrease in traffic, average time savings of more than 50% is achieved across different distance settings.
- It is observed that the alternatives involving emerging technology do not perform well for the best cost, time, and sustainable options. Although air taxis operate at a very high speed, there will be only very few operational sites within NYC at the initial stages. Nevertheless, in the long-run, with more infrastructure sites becoming operational, air taxis could serve as a potential middle-mile option, especially considering the time criteria.
- Despite the fact that robots are expected to save costs by replacing humans, they are limited by their speed, operating range, and capacity. Nonetheless, with better innovations in the future, robots could serve as a good cost-effective alternative.

## 6. Conclusions

Recent years have witnessed a significant increase in same-day delivery orders, specifically with the increasing presence of the online retail market. It is reported that over 50% of retailers offer same-day delivery services. Aside from large and traditional retailers that offer a vast amount of same-day delivery services, other non-traditional couriers can provide similar instant delivery services. Motivated by a real-life case study, this paper focuses on developing a simulation algorithm that assists these instant package delivery companies in providing the best solution with respect to three criteria: cost, time, and sustainability. For the purpose of illustrating our approach, we consider the delivery options in New York City. Based on our analysis, we propose several managerial implications. One of the major drawbacks of this study is that it assumes there will be only one delivery made at a time. As a potential scope of improvement, this assumption could be circumvented considering multiple package deliveries. Moreover, other emerging technologies, such as drones, could also be considered in future research.

## References

Amazon PrimeAir. (2020). Retrieved February 4, 2020, from https://www.amazon.com/Amazon-Prime-Air/b?ie=UTF8&node=8037720011

Angelovska, N. (2019). Top 5 Online Retailers: 'Electronics And Media' Is The Star Of E-commerce Worldwide. Retrieved February 4, 2020, from

https://www.forbes.com/sites/ninaangelovska/2019/05/20/top-5-online-retailers-electronics-and-media-is-the-star-of-e-commerce-worldwide/#36c33ebf1cd9

Bailey, S. (2020). Drones could help fight coronavirus by delivering medical supplies. Retrieved May 19, 2020, from https://edition.cnn.com/2020/04/28/tech/zipline-drones-coronavirus-spc-intl/index.html

Barclays. (2014). The Last Mile: Exploring the Online Purchasing and Delivery Journey.

Baron, Ethan (2019). Amazon looks to turn public buses into mobile delivery stations. https://www.mercurynews.com/2019/01/29/amazon-looks-to-turn-public-buses-into-mobile-delivery-stations.

Boyer, K. K., Prud' homme, A. M., & Chung, W. (2009). The last mile challenge: evaluating the effects of customer density and delivery window patterns. Journal of business logistics, 30(1), 185-201.

Burszytynsky (2020). https://www.cnbc.com/2020/04/02/coronavirus-instacart-is-giving-its-shoppers-health-and-safety-kids.html

Butler, S. (2020, March 10). Delivery and digital services thrive on coronavirus outbreak. Retrieved May 19, 2020, from https://www.theguardian.com/business/2020/mar/10/delivery-and-digital-services-thrive-on-coronavirus-outbreak.

Cheng, 2019: https://www.forbes.com/sites/andriacheng/2019/12/04/amid-record-online-sales-new-york-is-working-with-amazon-others-to-test-cargo-bike-delivery-your-city-will-be-next/#58e51179393f

Dayarian, I., Savelsbergh, M., & Clarke, J.-P. (2020). Same-Day Delivery with Drone Resupply. Transportation Science. doi: 10.1287/trsc.2019.0944.

Devari, A., Nikolaev, A. G., & He, Q. (2017). Crowdsourcing the last mile delivery of online orders by exploiting the social networks of retail store customers. Transportation Research Part E: Logistics and Transportation Review, 105, 105-122.

Dolan, S. (2018). The challenges of last mile delivery logistics & the technology solutions cutting costs. Retrieved February 4, 2020, from https://www.businessinsider.com/last-mile-delivery-shipping-explained

Environmental Health & Safety (2014). https://viceprovost.tufts.edu/ehs/files/October2014.VI2-Backpacks.pdf

FedEx (2020). https://about.van.fedex.com/newsroom/fedex-welcomes-roxo-the-fedex-sameday-bot-to-the-u-a-e/

Fuldauer, E. (2019, May 17). The Last Mile issue: how can we solve urban delivery problems? Retrieved February 4, 2020, from https://www.smartcitylab.com/blog/mobility/the-last-mile-issue-how-can-we-solve-urban-delivery-problems/

Gevaers, R., Van de Voorde, E., & Vanelslander, T. (2009). Characteristics of innovations in last-mile logistics-using best practices, case studies and making the link with green and sustainable logistics. Association for European Transport and contributors.

Glassdoor. (2021). https://www.glassdoor.com/Salaries/new-york-city-delivery-driver-salary-SRCH_IL.0,13_IM615_KO14,29.htm

Haag, M., & Hu, W. (2019). 1.5 Million Packages a Day: The Internet Brings Chaos to N.Y. Streets. Retrieved February 4, 2020, from https://www.nytimes.com/2019/10/27/nyregion/nyc-amazon-delivery.html

Hartman, N. (2018). Same Day Delivery Trends and Statistics to Guide Your Business. Retrieved February 4, 2020, from https://www.gopeople.com.au/blog/same-day-delivery-trends-and-statistics-to-guide-your-business/

Hausmann, L., Herrmann, N.-A., Krause, J., & Netzer, T. (2014). Same-day delivery: The next evolutionary step in parcel logistics. Retrieved February 4, 2020, from https://www.mckinsey.com/industries/travel-transport-and-logistics/our-insights/same-day-delivery-the-next-evolutionary-step-in-parcel-logistics33

Hu, W. (2018). New York Subway's On-Time Performance Hits New Low. The New York Times. https://www.nytimes.com/2018/03/19/nyregion/new-york-subways-on-time-performance-hits-new-low.html

Hu, W. (2020). https://www.nytimes.com/2020/04/09/nyregion/nyc-coronavirus-empty-streets.html

Kawa, A., Pieranski, B., & Zdrenka, W. (2018). Dynamic configuration of same-day delivery in E-commerce. In Modern Approaches for Intelligent Information and Database Systems (pp. 305-315). Springer, Cham.

Kim, E. (2019). Amazon can already ship to 72% of US population within a day, this map shows. Retrieved February 4, 2020, from https://www.cnbc.com/2019/05/05/amazon-can-already-ship-to-72percent-of-us-population-in-a-day-map-shows.html

Klapp, M. A., Erera, A. L., & Toriello, A. (2018). The Dynamic Dispatch Waves Problem for same-day delivery. European Journal of Operational Research, 271(2), 519–534. doi: 10.1016/j.ejor.2018.05.032.

Kung, L. C., & Zhong, G. Y. (2016). Platform delivery: a game-theoretic analysis of a new delivery model in the sharing economy. In Pacific Asia Conference on Information Systems (PACIS). Association for Information System.

Lin, J., Zhou, W., & Du, L. (2018). Is on-demand same day package delivery service green? Transportation Research Part D: Transport and Environment, 61, 118–139. doi: 10.1016/j.trd.2017.06.016.

Love, J., & Peters, K. T. (2006). Internet retailer magazine top 500 guide: 2006 edition. Chicago: Vertical WebMedia.

Mazur (2015). https://www.vox.com/2015/10/8/9480951/bike-commute-data-strava

Metropolitan Transportation Authority (2020). https://new.mta.info/coronavirus/subway-and-bus-service

New York City (2006). https://www1.nyc.gov/assets/planning/download/pdf/plans/transportation/td_pedloschapterfive.pdf

New York City Department of Transportation (2018). http://www.nyc.gov/html/dot/downloads/pdf/mobility-report-2018-print.pdf

Ni, M., He, Q., Liu, X., & Hampapur, A. (2019). Same-Day Delivery with Crowdshipping and Store Fulfillment in Daily Operations. Transportation Research Procedia, 38, 894–913. doi: 10.1016/j.trpro.2019.05.046.

NYC Department of City Planning (2020). https://www1.nyc.gov/site/planning/planning-level/nyc-population/population-facts.page.

Office Depot (2020). Retrieved May 19, 2020, from https://www.officedepot.com/cm/help/samedaydelivery

PricewaterhouseCoopers (2019). Retrieved February 4, 2020, from https://www.pwc.com/gx/en/industries/consumer-markets/consumer-insights-survey.html

Rajendran, S., & Shulman, J. (2020). Study of emerging air taxi network operation using discrete-event systems simulation approach. Journal of Air Transport Management, 87, 101857.

Rajendran, S., & Srinivas, S. (2020). Air taxi service for urban mobility: a critical review of recent developments, future challenges, and opportunities. Transportation research part E: logistics and transportation review, 143, 102090.

Rajendran, S., Srinivas, S., & Grimshaw, T. (2021). Predicting demand for air taxi urban aviation services using machine learning algorithms. Journal of Air Transport Management, 92, 102043.

Rajendran, S., Srinivas, S., & Saha, C. (2015). Analysis of operations of port using mathematical and simulation modelling. International Journal of Logistics Systems and Management, 20(3), 325-347.

Rougès, J., & Montreuil, B. (2016). Crowdsourcing delivery: New interconnected business models to reinvent delivery. 1st International Physical Internet Conference.

Salama, M., & Srinivas, S. (2020). Joint optimization of customer location clustering and drone-based routing for last-mile deliveries. Transportation Research Part C: Emerging Technologies, 114, 620-642.

Scherr, Y. O., Neumann-Saavedra, B. A., Hewitt, M., & Mattfeld, D. C. (2018). Service network design for same day delivery with mixed autonomous fleets. Transportation research procedia, 30, 23-32.

Seow, C., Delaney-Klinger, K., Boyer, K. K., & Frohlich, M. (2003). The return of online grocery shopping: a comparative analysis of Webvan and Tesco's operational methods. The TQM Magazine.

Sheth, M., Butrina, P., Goodchild, A., & Mccormack, E. (2019). Measuring delivery route cost trade-offs between electric-assist cargo bicycles and delivery trucks in dense urban areas. European Transport Research Review, 11(1). doi: 10.1186/s12544-019-0349-5

Slabinac, M. (2015). Innovative solutions for a "Last-Mile" delivery–a European experience. Business Logistics in Modern Management.

Smith, D., & Srinivas, S. (2019). A simulation-based evaluation of warehouse check-in strategies for improving inbound logistics operations. Simulation Modelling Practice and Theory, 94, 303-320.

Stonehem, B. (2016). UberEats Food Delivery: Learning the Basics.

Ulmer, M. (2017). Delivery deadlines in same-day delivery. Logistics Research, 10(3), 1-15.

Ulmer, M. W., & Thomas, B. W. (2018). Same-day delivery with heterogeneous fleets of drones and vehicles. Networks, 72(4), 475–505. doi: 10.1002/net.21855

Ulmer, M. W., & Thomas, B. W. (2018). Same-day delivery with heterogeneous fleets of drones and vehicles. Networks, 72(4), 475–505. doi: 10.1002/net.21855.

US Department of Transportation (2020) - https://www.transportation.gov/

Veeraraghavan, S., & Abdallah, T. (2019). https://knowledge.wharton.upenn.edu/article/amazons-shipping-challenges-will-out-of-the-box-solutions-work/

Voccia, S. A., Campbell, A. M., & Thomas, B. W. (2019). The Same-Day Delivery Problem for Online Purchases. Transportation Science, 53(1), 167–184. doi: 10.1287/trsc.2016.0732.

Xi, G., Cao, X., & Zhen, F. (2020). The impacts of same day delivery online shopping on local store shopping in Nanjing, China. Transportation Research Part A, 136, 1–13. doi: 10.1016/j.tra.2020.03.030.

Yao, B., McLean, C., & Yang, H. (2019). Robust optimization of dynamic route planning in same-day delivery networks with one-time observation of new demand. Networks, 73(4), 434-452.

Yazici, A., Ozguven, E. E., & Kocatepe, A. (2017). Urban Travel Time Variability in New York City: A Spatio-Temporal Analysis within Congestion Pricing Context.

Zhang, D., Ivanco, A., & Filipi, Z. (2015). An averaging approach to estimate urban traffic speed using large-scale origin-destination data. International Journal of Powertrains, 4(2), 126-140.